\documentclass{article}
\usepackage{amsfonts}

\usepackage{graphicx}
\usepackage{amsmath}

\input{tcilatex}

\begin{document}

\section{Theory of Electron Phenomena in Deformed Crystals}

Author: Ilya M. Dubrovskii, Institute for Metal Physics of National Academy
of Sciences of the Ukraine, 36, Academician Vernadsky Blvd UA-03680
Kyiv-142, Ukraine

\begin{center}
E-mail: lodub@dildub.kiev.ua\medskip
\end{center}

\subsubsection{Abstract}

On this article there are presented the main results of the theory of
electron phenomena in an unordered condensed matter that can be described as
an inhomogeneously deformed crystalline lattice with dislocations. The
one-electron effective Hamiltonian is derived by introducing a new basis for
the expansion of the sought functions. It allows to overcome the difficulty
of the discontinuity of the displacement field. The same method is used for
the derivation of the equations for eigentone of a deformed crystal, and the
equations of the theory of dynamical scattering of electrons by it. The
general problem of the description of the totality of electron states in a
deformed crystal is discussed, and some new solutions describing localized
electron or vibration states are obtained. There is derived and researched
the equation of an electrical field generated by the deformation in a
crystal with consideration for the interaction of electrons and ions by the
self-consisted field theory. The tensor kinetic coefficients are obtained,
which are proportional to the deformation tensor. Such kind of these
coefficients can explain some experimental effects. The appearing of the
local superconducting regions and the subsidiary solenoidal current that is
generated by a transport one in a deformed superconductor is considered. The
behavior and properties of the Abrikosov's vortex in anisotropic and/or
deformed superconductors are researched. In more detail this theory has been
published in the monograph by the same author of the same title in Russian.

\subsection{ 1. Account of Deformed Crystal}

In the title of this article as the ''deformed crystal'' the unordered
solid-state substance is named. In this crystal the equilibrium sites of
atoms form a lattice that can be described as a inhomogeneously deformed
crystalline lattice with dislocations. The electron potential energy differs
from the periodic function of coordinates in other way than in a crystal
disordered by localized defects, \textit{e. g., }interstitial or
substitutional impurities. In the site representation these cases can be
considered as alternative limiting cases, because the localized defects are
described by the diagonal disorder, but a deformation changes the
non-diagonal hamiltonian elements and makes them dependent on site
coordinates, and not only on their differences. Electron phenomena in
non-ideal crystals with localized defects have been investigated in a very
large number of experimental and theoretical works. Deformed crystals have
been investigated poorly, and the author devoted many years to their
systematic investigations based on the unified method of attack. This
article is an announcement of the main results of this research. In more
detail this theory has been published in the monograph [1] in Russian.

The proposed method of attack is applicable in the case, when the difference
of the deformed crystal from the ideal one can be described in the most part
of area by some small and smooth function of coordinates. We shall name
functions ''smooth'', when their variation on a distance of the order the
ideal crystal lattice constant is negligible. The other definition of the
smooth function: its Fourier-image is negligible outside of the vicinity of
the origin of coordinates of the reciprocal space with the radius much
smaller than the characteristic dimension of the Brillouin zone of the
corresponding ideal crystal. Functions that describe the crystal deformation
are introduced using Lagrange's coordinates. Let us conceive that before to
be deformed the ideal crystal lattice was rigidly connected with a
continuum, in which the Cartesian coordinate net of lines is embedded. When
the deformation is performed, the lines of this net are distorted and
transformed into the net of coordinate lines the Lagrange's curvilinear
coordinate system. The coordinates in this system and the Cartesian
coordinates that are independent on the deformation (they are named Euler's)
are connected by the following equation:

\ \ \ \ \ \ \ \ \ \ \ $\ \ \ \ \ \ \ \ \ \ \ \ \ \ \ \ \ \ \ \ \ \ \ \ \ \ \
\ \ \ \ \ \ \ \ \ \ \ \ \ \ \ \ \ \ \ \ \ \ \mathbf{x}=\mathbf{y+u}\left(
\mathbf{y}\right) .$ \ \ \ \ \ \ \ \ \ \ \ \ \ \ \ \ \ \ \ \ \ \ \ \ \ \ \ \
\ \ \ \ \ \ \ \ \ \ \ \ \ \ \ \ \ (1) \ \ \ \ \ \ \ \ Here $\mathbf{x}$ is
the Euler's coordinate, $\mathbf{y}$ is the Lagrange's coordinate, and $%
\mathbf{u}\left( \mathbf{y}\right) $ is the displacement field. In the cases
when the displacements are small and are unambiguously defined this field
can be used for the describing of the deformation. As this takes place, the
displacements can be considered as small parameters of the perturbation for
the hamiltonian of an electron in the periodic field. This approach is
commonly employed in the theory of the electron-phonon interaction, in which
deformation is generated by vibrations of atoms near the sites of the ideal
lattice. In the case of static deformation the displacements are usually
non-small, and when the deformation is generated by topological defects,
\textit{i. e. }by dislocations, they can't be described by a continuous
unambiguous field. Then for the investigation of the electron-deformation
interaction it is useful to describe the deformation by the field of the
distortion tensor $\mathbb{W}$ that is introduced by using the differentials
of the Euler's and Lagrange's coordinates:

$\ \ \ \ \ \ \ \ \ \ \ \ \ \ \ \ \ \ \ \ \ \ \ \ \ \ \ \ \ \ \ \ \ \ d%
\mathbf{x}=\left( \mathbb{I}+\mathbb{W}\right) d\mathbf{y},$ \ \ \ \ \textit{%
i. e. \ \ \ }$dx_{i}=\left( \delta _{ij}+w_{ij}\right) dy_{j}.$ \ \ \ \ \ \
\ \ \ \ \ \ \ \ \ \ \ \ \ \ \ \ \ \ \ \ \ \ \ \ \ This underlines the need
for the general case to take precisely the distortion tensor, but not its
symmetric term, \textit{i. e.} deformation, as it was made elsewhere. The
point is that the deformation describes the change of the shape of a cubic
volume element, and the antisymmetric term of the distortion tensor
describes its turning. It is obvious that, when the no-deformed crystal was
anisotropic, the local turning is also significant for the properties of the
deformed crystal. The distortion tensor is always an unambiguous field. It
is dimensionless, small and smooth in the most part of the crystal volume.
It can have singularities in some points and dislocation lines, and then
takes large values in the vicinities of these singularities. If these areas
would be eliminated from the consideration, the distortion tensor can be
taken as the useful phenomenological parameter.2. Effective Hamiltonian
Method for One-Electron Problem in Deformed Crystal

\subsubsection{\qquad2.1. Short History of Problem}

The fundamental method of studying electron states in a deformed crystal
that consists of neutral and short-range ions, which is used in the
monograph [1] and elsewhere, is the method of the effective hamiltonian. It
was first put forward by S. I. Pecar [2], and next was developed and
justified in many works, from which the works by J. M. Luttinger and W. Kohn
[3,4] are best suited to our purpose. The fundamental task of this method is
the construction of the operator, smooth eigenfunctions of which are some
eigenfunctions of the initial hamiltonian with averaged short-wave
variations, and the corresponding eigenvalues describe the part of the
energy spectrum. This effective hamiltonian usually involves the kinetic
energy and the smooth term of the potential energy of the initial
hamiltonian. The kinetic energy is the dependence of the energy on
quasi-momentum for the considered band (the law of dispersion). in which the
quasi-momentum is changed by the kinematic momentum operator. The vector
potential of the external magnetic field is included by the common manner in
the definition of the kinematic momentum. By this means the smooth external
fields superimpose into the effective hamiltonian without any change. In all
methods of derivation of the effective hamiltonian the translational
symmetry of the non-perturbed hamiltonian of the electron in the crystal is
essentially used.

Inasmuch as the deformation of the crystal can be described by a smooth
function, some authors [5 - 10] proposed to introduce an effective
hamiltonian of the electron that depends on the distortion by the
phenomenological consideration. In so doing as a rule some unjustified
assumption are made. Every so often with an assumption that the kinematic
momentum is small there isn't taken into account the dependence of the
kinetic energy on the distortion, and the deformation is supposed to
generate only the deformation potential, which is proportional to
distortion. It is incorrect, because the terms of the hamiltonian that
depend on the momentum operator can make a radical change of the spectrum,
even though their multipliers are small. Example is provided by the change
of the electron states by a weak magnetic field. In the case of a shear
deformation in a cubic crystal the deformation potential of the first order
is absent by symmetry. Then in the work [7] the amendment proportional to
the distortion tensor was inserted into the reciprocal effective mass, but
proportionality factor was supposed to be a scalar. In all cited works the
introduced effective hamiltonian was used for the solving of the special
tasks, and its general characteristics weren't researched.

A great number of works are known, in which the effective wave equation in
the approximation of the effective mass has been deduced for the simple
model of an electron in the potential field generated by the deformed cubic
lattice of short-range ions. This derivation runs into the problem that the
difference of the potential energies of the electron in the deformed crystal
is a non-local, non-small, and non-smooth function of coordinates. Let us
suppose that the potential energy of the electron has the form:

$\ \ \ \ \ \ \ \ \ \ \ \ \ \ \ \ \ \ \ \ \ \ \ \ \ \ \ \ \ \ \ \ \ \ \ \ \ \
\ \ \ \ \ V\left( \mathbf{x}\right) =$ $\sum\limits_{s}V_{a}\left( \mathbf{%
x-X}_{s},\mathbb{W}\left( \mathbf{x}\right) \right) ,$ \ \ \ \ \ \ \ \ \ \ \
\ \ \ \ \ \ \ \ \ \ \ \ \ \ \ \ \ \ \ \ \ \ \ \ \ \ \ \ \ \ \ \ \ \ \ \
where \textbf{X}$_{s},$\textbf{x }are the coordinates of the atoms and the
electron in the Euler's system of axes, and the function $V_{a}$ declines
rapidly when the first argument increases. It is easily seen that this model
incorporates the commonly used models of hard ions and deformable ones. The
function $V\left( \mathbf{x}\right) $ can be represented as the sum of the
periodic function and the terms, which are proportional to the powers of the
distortion components only if the transformation to the embedded Lagrange's
coordinates (1) would be made. If the lattice contains dislocations, that
coordinates are discontinuous functions of the Euler's ones, because the
displacement field generated by a dislocation has a jump equal to the
Burgers vector of this dislocation at some surface bounded by the
dislocation line. In other respects the choice of this surface is voluntary,
therefore the choice of the Lagrange's coordinates isn't defined unique. It
can be made unique if the requirement would be taken that the quantity of
the material isn't changed with deformation. The potential energy is a
continuous function of the Lagrange's coordinates, because the distortion
components are continuous functions,with the exception of the dislocation
lines, and their multipliers prove to be periodic functions, which have the
full translational symmetry of the ideal lattice. Since the Burgers vector
is always one of the vectors of the ideal lattice (a splitted dislocation
can be considered as the dislocation with the extended core), then the
argument of the periodic function change of the Burgers vector doesn't vary
the value of this function. The displacement are not defined on the
dislocation line, and the distortion components indefinitely increase, as
this line is approached, therefore the transformation to the Lagrange's
coordinates and the expansion in terms of the distortion components can't be
made in the vicinity of dislocations.

All works use the transformation of the coordinates (1) in one form or
another. As a rule, in them there was not taken into account the fact that
the area of applicability of this expansion of the potential energy is
multiply connected. The other cause of mistakes in the effective hamiltonian
derivation in the case of a crystal with dislocations is the neglect of the
discontinuity of the transformation (1). If this transformation is used at
once to the one-electron Schr\"{o}dinger equation, as done in the works [11
- 13], then for the continuity of the wave function in the Euler's
coordinates special boundary conditions must be defined on the cut surfaces.
It makes impossible using the customary method of derivation. The derivation
method, which is used in the works [14, 15], can be reduced to the
computation of the matrix elements of the one-electron hamiltonian in the
representation, the basis of which is constituted by the eigenfunctions for
the ideal crystal (the Bloch functions) in which $\mathbf{y}\left( \mathbf{x}%
\right) $ is substituted instead the coordinates $\mathbf{x}$. Using this
basis is researched in more detail in the work [16]. In the case of the
crystal with dislocations its basis vectors are discontinuous functions,
therefore the expansion of the wave functions over them is impossible. These
methods can be used only in the case of the compatible deformation. In
particular, the transformation of the coordinates (1) is used for the
computation of the energy spectrum in uniformly deformed semiconductors in a
number of works summarized in the monograph [17]. However, the extrapolation
of these results to the case of non-uniform deformation used in the work
[18] is valid only, if the characteristic distance of the deformation
variation is considerably greater than the quantity reciprocal to the
quasi-momentum modulus. Other attempts to deduce the effective hamiltonian
were founded on the representation, the basis of which was constructed from
modified Wannier functions [19 - 23]. In these works the fact of small
variation of distortion at the distance of the atomic potential influence or
of the radius of the Wannier functions was used. This idea was most
thoroughly investigated in the work [23]. In so doing it turned out that the
diagonalization with respect to the band number with obtaining the one-band
effective hamiltonian is impossible.

\subsubsection{\qquad2.2 . Phenomenological Theory of Effective Hamiltonian}

In the section [1.2.3] (here and in what follows in references to the
monograph [1] we shall indicate the part number by the second figure, and
the section by the third one) the most general effective hamiltonian with
the quadratic dispersion law was considered:

\ \ $\widehat{H^{\alpha}}=\frac{1}{2}\widehat{p_{i}}\mu_{ij}^{\alpha}\left(
\delta_{jk}+2A_{jk}^{\alpha}\right) \widehat{p_{k}}+V_{1}+V_{2}+V_{3},$ \ \
\ $A_{kl}^{\alpha}=\frac{1}{2}T_{ijkl}^{\alpha}w_{ij}.$ \ \ \ \ \ \ \ \ \ \
\ \ \ \ \ \ \ \ \ \ \ \ \ \ \ (2) \ \ \ \ \ \ \ \ \ \ \ \ \ \ \ \ \ \ \ \ \
\ \ Here\ $\mu_{ij}^{\alpha}$ are the components of the tensor of the
reciprocal effective mass for the band $\alpha$ (in what follows the band
index will be omitted). There are taken into account the potential terms not
only of the first order $V_{1}$, but also of the second order $V_{2}$, and
of the third order $V_{3},$ because for the quadratic dispersion law not
only the distortion components, but also the momentum components must be
consider as small, and therefore the terms of the third order were
considered in the operator of the kinetic energy. This operator can be
equivalently rewritten as the quadratic form in new kinematic momentum

\bigskip \ $\ \ \ \ \ \ \ \ \ \ \ \ \ \ \ \ \ \ \ \ \ \ \ \ \ \ \ \ \ \ \ \
\ \widehat {p_{k}}=-i\hbar\left( \frac{\partial}{\partial x_{k}}+A_{jk}\frac{%
\partial }{\partial x_{j}}+\frac{1}{2}\frac{\partial A_{jk}}{\partial x_{j}}%
\right) $ \ \ \ \ \ \ \ \ \ \ \ \ \ \ \ \ \ \ \ \ \ \ \ \ \ \ \ \ \ \ \ \ \
\ \ \ \ \ \ \ \ \ \ \ \ \ \ \ \ \ \ \ \ \ \ \ \ \ \ \ \ \ \ \ \ \ with the
constant coefficients $\mu_{ij}^{\alpha}.$ With using this determination of
the kinematic momentum the phenomenological effective hamiltonian can be
generalized to the case of the arbitrary dispersion law. In addition it
permits to reveal the principal difference between compatible and
incompatible deformations. If the tensor $\mathbb{A}$ satisfies the
compatibility condition, \textit{i. e. }if it can be represented\ as the
gradient of a continuous and unambiguous vector field $\mathbf{u}^{\prime}$,
then by the change of the sought function $\psi=\exp\left( \frac{A_{ll}}{2}%
\right) \varphi,$ and by the transformation of coordinates $\mathbf{x}=%
\mathbf{y-u}^{\prime}$ the subsidiary terms in the definition of the
kinematic momentum can be eliminated. Then the problem is reduced to the
common effective equation with a potential. This is like to the elimination
of the vector potential of the magnetic field that has zero curl, other than
only phase but also modulus of the wave function as well as the coordinates
change. If \ $\mathbb{A}$ is incompatible tensor $\mathbf{u}^{\prime}$ can't
be introduced as a continuous field, and therefore the coordinate change
can't be made. Then the solving of the equation with variable multipliers at
the derivatives is indispensable. It should be noted that the compatibility
of the deformation not necessarily ensures the compatibility of the tensor $%
\mathbb{A}$.

\subsubsection{\qquad2.3. Derivation of Effective Hamiltonian by Introducing
of Deformation Basis}

In the section [1.2.4] the hamiltonian, which is similar to the
phenomenological one, was derived for the simple model described by the
hamiltonian of an electron in the potential field generated by the deformed
cubic lattice of short-range ions. There was considered the non-degenerated
band that has the minimum of energy in the center of the Brillouin zone.
With this derivation the limits of the applicability of this hamiltonian
were determined and the interpretation of the effective wave function was
specified.

The fundamental idea, which permits in the section [1.2.4] to overcome the
pointed difficulties in the derivation of the effective hamiltonian in the
case of the crystal containing dislocation, is to introduce the new basis in
the Hilbert's space of the wave functions, which we shall name the
deformation basis.

Let us present shortly the consideration about introducing this basis. The
plane waves that can be the basis system of the vectors in the Hilbert's
space can be rewritten in the form of the functions, which have the property
of Bloch functions for the considered lattice.

$\ \ \ \ \ \ \ \ \ \ \ \ \Omega^{-\frac{1}{2}}\exp\left( i\mathbf{Kx}\right)
=\Omega^{-\frac{1}{2}}\exp\left( i\mathbf{qx}\right) \exp\left( i\mathbf{gx}%
\right) =\left| \mathbf{qg}\right\rangle _{B}.$ \ \ \ \ \ \ \ \ \ \ \ \ \ \
\ \ \ \ \ \ \ \ \ \ \ \ \ \ \ \ \ \ \ \ \ \ \ \ (3) \ \ \ \ \ \ \ \ \ \ \ \
\ \ \ \ \ \ Here $\mathbf{q}$ is situated in the first Brillouin zone, and $%
\mathbf{g}$ is a reciprocal lattice vector of the crystal under
consideration. Let us name as the deformation basis the set of functions

$\ \ \ \ \ \ \ \ \ \ \ \ \ \ \ \ \ \ \ \left| \mathbf{qg}\right\rangle
_{D}=\Omega^{-\frac{1}{2}}\exp\left( i\mathbf{qx}\right) \exp\left( i\mathbf{%
gy}\left( \mathbf{x}\right) \right) ,$ \ \ \ \ \ \ \ \ \ \ \ \ \ \ \ \ \ \ \
\ \ \ \ \ \ \ \ \ \ \ \ \ \ \ \ \ \ \ \ \ \ \ \ \ \ \ \ \ \ \ \ \ \ (4) \ \
\ \ \ \ \ \ \ \ \ $\mathbf{y}\left( \mathbf{x}\right) $ being defined from
(1). Therefore $\mathbf{y}\left( \mathbf{x}\right) $ has the jumps by the
Burgers vector. These jumps, when they are multiplied by the reciprocal
lattice vector, vary the exponent of power by $2\pi ni,$ therefore the
functions of the deformation basis are continuous and defined elsewhere
except the dislocation lines that construct in the three-dimensional space a
set of measure zero. It can be shown that, if the modulus of $\mathbf{q}$ is
considerably smaller than the Brillouin zone dimension, the deformation
basis functions with different $\mathbf{q}$ and/or $\mathbf{g}$ are
approximately orthogonal. The set of the one-electron Bloch eigenfunctions
of the hamiltonian in the ideal crystal would be got from the set of the
Bloch plane waves (3) by using the unitary matrix $\left\| B\right\| ,$
which is the direct sum over all $\mathbf{q}$ of the unitary matrices $%
\left\| B\left( \mathbf{q}\right) \right\| .$ These matrices connect the
Bloch plane waves (3) having coincident $\mathbf{q}$ with the eigenfunctions%
\textbf{\ }in\textbf{\ }the crystal\textbf{, }for which this $\mathbf{q}$ is
the quasi-momentum:

$\ \ \ \ \ \ \ \ \ \ \ \ \ \ \ \ \ \ \ \ \ \ \ \ \ \ \ \ \ \ \ \ \ \ \ \ \ \
\ \ \ \ \ \ \ \ \ \ \Psi _{\alpha\mathbf{q}}\left( \mathbf{x}\right)
=\sum\limits_{\mathbf{g}}\ \left| \mathbf{qg}\right\rangle _{B}B_{\alpha%
\mathbf{g}}\left( \mathbf{q}\right) .$ \ \ \ \ \ \ \ \ \ \ \ \ \ \ \ \ \ \ \
\ \ \ \ \ \ \ \ \ \ \ \ \ \ \ \ \ \ \ \ \ \ \ \ \ \ \ \ \ \ \ \ In Luttinger
- Kohn method of the derivation of the effective hamiltonian near to the
selected point of the Brillouin zone $\mathbf{q}_{0}$ the basis of the
representation is got from the Bloch plane waves by the unitary matrix,
which is constructed by the same direct sum of the coincident unitary
matrices $\left\| B\left( \mathbf{q}_{0}\right) \right\| .$ Then in the
matrix of the hamiltonian of the electron in the ideal crystal and in a
smooth external field the block of the elements with small $\mathbf{k=q-q}%
_{0}$ and $\mathbf{k}^{\prime}\mathbf{=q}^{\prime}\mathbf{-q}_{0}$ can be
found, for which the interband elements $H_{\mathbf{k\alpha,k}^{\prime}\beta}
$ are small and can be eliminated by ''$\mathbf{k}\cdot\mathbf{p"}$
perturbation theory. The remaining problem of the diagonalization of the
block relating to the single band with respect to the vector indices $%
\mathbf{k,k}^{\prime}$ can be reduced to the problem of the eigenvalues for
the differential operator of the second order that is the effective
hamiltonian in the effective mass approximation. Its eigenvalues, which
correspond to the smooth eigenfunctions, in sum with $E\left( \mathbf{q}%
_{0}\right) $ present some part of the energy spectrum of the real
hamiltonian, and corresponding wave functions are approximately described by
the products of the effective eigenfunctions and $\Psi_{\alpha\mathbf{q}%
_{0}}\left( \mathbf{x}\right) .$

When the effective hamiltonian for the electron in the deformed crystal is
derived, the basis for the expansion of the sought function is formed
similarly, instead of the Bloch plane waves (3) the functions of the
deformation basis (4) are used. The expansion over this basis can determine
the wave function only in the multiply connected area, from which the
dislocation lines are eliminated. Because in the vicinities of these lines
the distortion components are very large, these vicinities must be also
eliminated from the consideration, and the boundary conditions for the
eigenvalue problem on the surfaces of these ''dislocation cores'' must be
introduced. The power series of the ''$\mathbf{k}\cdot \mathbf{p"}$
perturbation theory must be taken in the third order, and then the effective
hamiltonian would have the form (2), if the ideal crystal dispersion law has
the extremum \ in the point $\mathbf{q}_{0}.$ In so doing the
phenomenological constants that are the tensor $\mathbb{T}$ components and
the coefficients defining the potential terms are expressed in terms of the
matrix elements of certain operators in the representation, which has as the
basis the Bloch functions of the ideal crystal. The main result of the
derivation of the effective hamiltonian from this model is the elucidation
of the facts: the effective equation must be solved with the boundary
conditions on the surfaces of dislocation cores, and its solution is the
multiplier at the Bloch eigenfunction in the ideal crystal in the band under
consideration with quasi-momentum $\mathbf{q}_{0},$ in the periodical factor
\ of which $\mathbf{x}$ is changed for $\mathbf{y}\left( \mathbf{x}\right) .$
The effective hamiltonian has meaning only for a small vicinity of the
selected point $\mathbf{q}_{0},$ \textit{i. e. }only smooth eigenfunctions
of it that vary no faster than the deformation are of physical sense. If in
the point $\mathbf{q}_{0}$ the ideal crystal dispersion law has an extremum,
then the effective hamiltonian involves only the terms quadratic in the
momentum components. The coefficients of these terms are the components of
the tensor of the reciprocal effective mass , and depend on coordinates. In
a generally case the term linear in momentum will be also involved, it is
similar to presence of an external magnetic field that depends on $\mathbf{q}%
_{0}.$ In the case, when $N$ bands have in the point $\mathbf{q}_{0}$ the
coincident energies, the effective hamiltonian isn't similar to the particle
hamiltonian. It is the matrix of the rank $N$, the elements of which are the
operators, and its eigenvector involves $N$ components-functions. The
electron wave function is the sum of the $N$ \ products of these components
with the Bloch functions of the corresponding bands, in the periodic
multipliers of which $\mathbf{x}$ is changed for $\mathbf{y}\left( \mathbf{x}%
\right) .$2.4. Using of Deformation Basis in Problems of Dynamical
Scattering of Electrons and Small Vibrations of Deformed Crystal

It can be shown (see the section [1.2.6]) that the fundamental equations of
the theory of the dynamical scattering of electrons in an ideal crystal in
the two-rays approximation (see the monograph [24]) can be derived as the
effective wave equation for the reciprocal space point that is situated on
the Brillouin zone boundary. That can be made by some modification of the
Luttinger - Kohn method. The equations of the scattering by a deformed
crystal can be derived through the use of the deformation basis functions
instead the plane waves in this modified method. They coincide with the well
known Takagi's equations (see the work [25]), which have been derived from
the less rigorous considerations, but their solutions describe the
amplitudes of the deformation basis functions rather than of the plane waves.

The problem of the eigentones of a deformed crystal that is considered in
the section [1.2.5] closely resembles the electron problem described above.
Let us choose the displacements of atoms from the sites of the deformed
lattice as coordinates, on which the potential energy of the crystal
depends. Then zero displacements correspond to the minimum of the potential
energy, though not absolute one. The set of motion equations is got by the
power series expansion of the potential energy in the second order, and
describes small vibrations of atoms about the equilibrium positions that
constitute the deformed lattice. Let as search the solution of this set in
the form:

$\ \ \ \ \ \ \ \ \ \ \ \ \ \ \ \ \ \ \ \ \mathbf{u}^{\nu}\left( \mathbf{X}%
_{0}^{\nu},t\right) =\mathbf{U}\left( \mathbf{X}_{0}^{\nu}\right) \exp\left(
i\mathbf{kX}_{0}^{\nu}\right) \left( 1+\frac{1}{2}w_{jj}^{\nu }\right)
\cos\omega t.$ \ \ \ \ \ \ \ \ \ \ \ \ \ \ \ \ \ \ \ \ \ \ \ \ \ \ \ \ \ \ \
\ \ \ \ \ \ \ \ \ \ \ \ \ \ \ \ \ \ \ \ \ \ \ \ \ \ \ Here the Greek indices
number the lattice sites, $\mathbf{X}_{0}^{\nu}$ are the coordinates of
these displaced sites. It can be seen that this solution differs from the
common wave solution in the perfect crystal by the dependence on the site
coordinates of the vector describing the magnitude and polarization of
vibrations. This is true, because the translation nonexists that retains the
lattice transferring its sites one into other. If it is granted that the
values of \ $\mathbf{U}\left( \mathbf{X}_{0}^{\nu}\right) $ little differ in
the neighboring sites, this vector can be described by a smooth function of
the continual argument. Then the set of the three differential equations,%
\textit{\ i. e. }the\textit{\ }matrix eigenvalue problem, for the components
of this vector can be obtained. These eigenvalues define the squared
eigentone frequencies, and three frequencies correspond to any wave vector.
The corresponding eigentones are the plane waves with modulated magnitudes
and polarizations. In an ideal crystal in the symmetrical point of the
Brillouin zone the vibrations can be presented as the longitudinal and two
transverse waves with coinciding frequencies of the transverse ones. Then in
the deformed crystal also one equation can be separated out of the set,
which describes the mainly longitudinal vibrations, which have weakly
modulated polarization. In transverse waves the vibration polarizations
would strongly depend on coordinates, and the space beatings would take
place. The magnitude modulation can be deep, \textit{i. e.} the local, but
sufficiently spacious can exist.

\subsection{3. Electron States in Deformed Crystal}

\subsubsection{\qquad3.1. General Consideration about Electron States in
Deformed Crystal}

The effective one-electron hamiltonian derivation is only the first
necessary step to the creation of the electron phenomena theory. In
principle, the full set of the eigenstates and the energy spectrum of the
electron in the band can be obtained by selecting some set of points in the
Brilloin zone of the non-deformed crystal, deriving the effective
hamiltonian in these points, and solving the eigenvalue problems. Each
eigenstate obtained by this way will be characterized by not only the band
number (if the band in this point is non-degenerated in the ideal crystal)
and the energy, but also by the Brillouin zone point, in which it has been
obtained. It is clear that in so doing this point isn't an exact quantum
number. But the making of this description demands exact knowledge of the
distortion tensor and the boundary conditions on the dislocation core
surfaces. In the most interesting cases the distortion field and the
boundary conditions can be neither well measured nor reproduced in the
predetermined view. Therefore only the observable quantities that are
defined by the statistical characteristics of the ensemble of possible
values of the distortion and boundary conditions have the physical sense.
The theory of this type was made, \textit{e. g.,} for the kinematic
scattering of X-rays in the monograph [26]. Some results in the electron
theory with taking into account only the deformation potential was obtained
by I. M. Lifshitz (see the review [27]). The dependence the kinetic energy
parameters (\textit{e. g. }the effective mass tensor components) on
coordinates can lead in addition to transformations in the one-electron
state system that are not evident at present time. The Hamilton's equations
that can be derived from the effective hamiltonian quadratic in the momentum
components by the reversion of the correspondence principle are non-linear,
because the derivatives of the classic hamiltonian with respect to
coordinates are the quadratic functions of the momentum components. In most
cases they are very complicated and supposedly non-integrable. As of now the
question about characteristics of the quantum system that corresponds to the
classical one, the motion of which is chaotic in the considerable proportion
of the phase volume, is insufficiently studied.

\subsubsection{\qquad3.2. Localized States in Potential Well Close to
Rectlinear Edge Dislocation}

Some cases of the electron states in defined distortion field have been
considered elsewhere and are described in the third chapter of the monograph
[1]. A number of works was devoted to localized electron states in the field
of the deformation potential that is proportional to the dilatation
generated by the rectilinear edge dislocation in an isotropic medium. In the
work [28] it is shown that the state energies, which are close to the
continuous spectrum, are approximately inversely proportional to their
number in the sequence arranged in the order of increasing. In the works
[29], and [30] the energy of the ground state was obtained by the
variational method. In the section [1.3.2] these results are improved by the
proposed modification of the variational method. The considered hamiltonian
is represented as the sum of two terms $\widehat{H}=\widehat{H_{0}}+\widehat{%
H_{1}}$ so that eigenfunctions of \ $\widehat{H_{0}}$ can be found. These
eigenfunctions are used as a trial functions in the standard variational
procedure. This decomposition is defined by some parameters that are
calculated further through the minimization of the expectation value\ $%
\widehat{H}$ in the ground state of $\widehat{H_{0}}$ . This expectation
value has the form of the expression of the ground state energy \ $\widehat{H%
}$ correct in the first order of the perturbation theory, in which $\widehat{%
H_{1}}$ is considered as a perturbation. There can be calculated the values
of the expansion parameters of this perturbation theory $\frac{\left\langle
0\right| \widehat{H_{1}}\left| m\right\rangle }{E_{0}-E_{m}},$ where $%
\left\langle 0\right| ,\left| m\right\rangle ,E_{0},E_{m}$ are the
eigenfunctions and corresponding eigenvalues of the operator \ $\widehat{%
H_{0}}$ with the values of the decomposition parameters obtained by
minimization $\left\langle 0\right| \widehat{H}\left| 0\right\rangle .$ If
these parameters of the perturbation theory are smaller than unity, the
obtained value of the ground-state energy can be improved by calculating the
amendment of the second order, which is always negative. The function of the
zero or the first order can be considered as the approximate expression for
the ground-state eigenfunction, and the energies and eigenfunctions of some
low-lying exited states can be calculated. It is apparent that this method
is inapplicable, when the distance between neighboring eigenvalues of the
hamiltonian $\widehat{H_{0}}$ is small. It would be, in particular, close to
the bound between discrete and continuous parts of the spectrum. When the
problem electron ground-state in the field of the deformation potential of
the edge dislocation is considered, it is found that the requirement of the
eigenfunction continuity leads to the necessity of its going to zero on the
dislocation line. This is an acceptable boundary condition for the effective
wave function, because then the indeterminacy of the real wave function on
the dislocation line disappears. But the extrapolation of the deformation
potential to the vicinity of the dislocation line is obviously incorrect,
because it is not only very large in thus vicinity, but also is
indeterminate on the dislocation line.3.3. Integrable Schr\"{o}dinger
Equation with Effective Mass Depending on Coordinates

Of particular interest is the effective wave equation in the case of
distortion field generated by a rectilinear screw dislocation in an
isotropic medium. The deformation potential of the first order in distortion
in this case is absent by symmetry and the effective mass depending on
coordinates has a dominant role. As discussed earlier precisely this kind of
perturbation is specific for the deformation effect. In this case the
variables in the equation can be disjointed in cylindric coordinate system.
It permits to integrate the equation completely, and to study the obtained
solutions. That was made partially in the works [28], [29]. Inasmuch the
analysis of this equation enables to consider some new characteristics of
the effective wave equation, let us dwell on it in detail.

In this case the hamiltonian (2) leads to the equation

$\ \ \ \ \ \ \ \ \ \ -\frac{\hbar^{2}}{2}\mu\left[ \nabla^{2}+\frac{b}{%
\pi\left( x_{1}^{2}+x_{2}^{2}\right) }T\left( -x_{2}\frac{\partial }{%
\partial x_{1}}+x_{1}\frac{\partial}{\partial x_{2}}\right) \frac{\partial }{%
\partial x_{3}}\right] \Psi+\frac{\upsilon}{x_{1}^{2}+x_{2}^{2}}\Psi=\left(
E-E_{\alpha}\right) \Psi.$ \ \ \ \ Here $\mu$ is the reciprocal effective
mass and $E_{\alpha}$ is the energy in the center of the Brilloin zone in
the band $\alpha,$ the constant $T$ is the shear component of the tensor $%
\mathbb{T},$ which is supposed to have the symmetry of \ isotropic medium, $%
\upsilon$ is the constant of the deformation potential of the second order,
the deformation potentials of the first and the third orders are zero by
symmetry. Let us divide this equation by $\hbar^{2}\mu,$ transform it to the
cylindrical coordinates and substitute $\Psi=\exp\left( ik_{3}z\right)
U\left( \rho,\varphi\right) .$ We obtain the equation:

$\ \ \ \ \ \ \ \ \ \ \ \ \ \ \ \ \ \ \ \ \ \ \ \ \ \ \ \ -\frac{1}{2}\left(
\frac{\partial^{2}U}{\partial\rho^{2}}+\frac{1}{\rho}\frac{\partial U}{%
\partial\rho}+\frac{1}{\rho^{2}}\frac{\partial^{2}U}{\partial\varphi^{2}}%
\right) -\frac{i\zeta}{\rho^{2}}\frac{\partial U}{\partial\varphi}+\frac
{\xi}{\rho^{2}}U=\sigma_{1}\frac{\varepsilon^{2}}{2}U.$ \ \ \ \ \ \ \ \ \ \
\ \ \ \ \ \ \ \ \ \ Here notes are introduced:

$\ \ \ \ \ \ \ \ \ \ \frac{bTk_{3}}{\pi}=\zeta,$ \ \ \ \ $\frac{\upsilon }{%
\hbar^{2}\mu}=\xi,$ \ \ sign$\left( E-E_{\alpha}\right) =\sigma_{1},$ \ \ $%
\left| E-\frac{\hbar^{2}}{2}\mu k_{3}^{2}-E_{\alpha}\right| =\frac{\hbar^{2}%
}{2}\mu\varepsilon^{2}.$ \ \ \ \ \ \ \ \ \ \ \ \ \ \ \ \ \ \ \ \ This
equation describes also the amplitude of eigentones in a body centered
crystal with the screw dislocation directed along the axis $\left[ 111\right]
,$ when its polarization is directed mainly parallel to the dislocation and
also the wave vector. It is significant that $\zeta$ can have positive or
negative values depending on to be aligned the Burgers vector either with
the quasi-momentum component $k_{3}$ or opposing it. This component is a
good quantum number, and can assume any values within the boundaries of the
Brillouin zone, because the crystal with the rectilinear dislocation retains
the translational symmetry in the dislocation direction. Then the equation
can be generalized to considering it as the Schr\"{o}dinger equation for the
hamiltonian, which is got by expansion of any dispersion law in terms of
small transversal components of quasi-momentum with any determined $k_{3}.$
In so doing in the determination of $\zeta$ instead of the multiplier $k_{3}$
an odd function of $k_{3}$ appears, which can be non-monotonous in
particular.

The solution of this equation have the form:

$\ \ \ \ \ \ \ \ \ \ \ \ \ \ \ \ \ \ \ \ \ \ \ \ \ \ \ \ \ \ \ \ \ \ \ \ \ \
\ \ \ \ \ \ \ U=\exp \left( im\varphi\right) Z_{\tau}\left( \rho\varepsilon%
\sqrt{\sigma_{1}}\right) ,$ \ \ \ \ \ \ \ \ \ \ \ \ \ \ \ \ \ \ \ \ \ \ \ \
\ \ \ \ \ \ \ \ \ \ \ \ \ \ \ \ \ \ \ \ \ \ \ \ \ \ \ \ \ \ \ where $m$ is a
natural number, $Z_{\tau}\left( z\right) $ is a cylindrical function with
the index

$\ \ \ \ \ \ \ \ \ \ \ \ \ \ \ \ \ \ \ \ \ \ \tau=\sqrt{\sigma_{2}\left|
m^{2}+2\zeta m+2\xi\right| },$ \ \ $\sigma_{2}=$sign$\left( m^{2}+2\zeta
m+2\xi\right) .$ \ \ \ \ \ \ \ \ \ \ \ \ \ \ \ \ \ \ \ \ \ \ \ \ \ \ \ \ \ \
\ \ \ \ \ \ \ \ \ \ \ \ \ \ \ \ \ \ \ \ \ \ \ The boundary condition on the
dislocation core in the works [28], [29] is supposed in the form of the
demand of the continuity of the solutions in the total space. For the
fulfillment of this condition there is supposed $\sigma _{1}=\sigma_{2}=1.$
In [1.3.3] the logarithmic derivative of the solution on the dislocation
core bound can assume any real values depending on the form of the wave
function in the dislocation core including the infinite value. Then with the
values $\sigma_{1}=1,$ \ $\sigma_{2}=-1$ and $\sigma_{1}=-1,$ \ $%
\sigma_{2}=-1$ the boundary condition can also be fulfilled. The equation
has been researched by the version of the JWKB method proposed in the
monograph [30]. This allows considering the effect of the boundary condition
and the correspondence between the characteristics of the equation solutions
and of the motion of the classical particle, the hamiltonian of which is
obtained without using the correspondence principle, which is non-founded
for the effective hamiltonian, and is inapplicable in the case of the
crystal vibrations. In both cases, when $\sigma_{1}=1,$ the solutions belong
to the continuous spectrum. By them the function describing the scattering
of an electron by the screw dislocation can be constructed. It was made in
the works [28], [29] as well, but in the section [1.3.3] there is first
pointed that, if for some values of $k_{3}$ would be $\sigma_{2}=-1,$ the
scattering by attraction replaces scattering by repulsion, and the
dislocation core must play a great part. The other essential singularity,
which also is not be noted elsewhere, is that the scattering by the screw
dislocation is non-symmetrical with respect to the plane, in which the
dislocation line and the electron quasi-momentum are situated.

\subsubsection{\qquad3.4. ''Kinetic Well'' in Classic and Quantum Problems}

When $\sigma_{1}=-1,$ \ $\sigma_{2}=-1$, the energy of the motion in the
transversal plane takes only discrete negative values, and the corresponding
states are localized. This is an example of the new type of the localization
that can originate from a special dependence of the effective mass on
coordinates. Unlike a classical particle, a band electron can have negative
kinetic energy (a hole), if the dispersion law has a maximum in the
Brillouin zone center. In the deformed crystal, when the dispersion law
depends on coordinates, the sign of the kinetic energy may turn out
different in various areas of the crystal. The electron, which has the
negative energy and non-zero momentum, moves, but it can't go out to the
area, in which its kinetic energy shall be positive. We can say that it is
situated in a ''kinetic well'' distinguished from the ordinary potential
one. In a similar manner it may be, when the summand of the kinetic energy,
which is an independent constant of motion, has different sign in various
areas of the crystal. In the section [1.3.3] there is obtained the
dependence of the discrete eigenvalues arranged in order of increasing on
their number. Close to the bound between discrete and continuous spectra
their absolute values constitute a decreasing geometric progression.

When the momentum component operators in the effective hamiltonian are
changed for $c-$quantities, the quasi-hamiltonian of the classic particle
would be obtained. It can't be contended that the phase trajectory of this
particle describes the variation with time of the averaged characteristics
of the electron wave packet. But there is no question that the properties of
the Hamilton's equations are connected with the properties of the
Schr\"{o}dinger equation, although this connection isn't studied adequately.
In section [1.3.4] the motion of this classic particle is considered in the
case of a screw dislocation with the condition of mirror-like reflection
from the cylindric dislocation core. The motion equations in this case can
be exactly solved. At some value of the hamiltonian parameters there is an
area of the phase space in which the motion in the plane perpendicular to
the dislocation is finite and conventionally-periodical. By using Bohr -
Sommerfeld quantization rules the same spectrum of the quasi-classical
energy levels can be obtained. When a small regular perturbation is added
into the hamiltonian, an area of a chaotic motion arises in the vicinity of
the separatrix that delimits the areas of finite and infinite motions. What
are the characteristics of the quantum system that correspond to this
feature of its classical analogue is not clarified presently.

An other example of the kinetic well is considered by using the classical
analogue in the section [1.3.5]. If an edge rectilinear dislocation in an
elastically isotropic crystal with the Poisson constant equal 0.5 is
considered, the deformation potential in the effective hamiltonian is equal
to zero. The disturbance of the deformation and the change of the lattice
topology in this case are described by the dependence of the effective mass
on coordinates. The solving of \ Schr\"{o}dinger equation in this case has
not met with success, but the corresponding Hamilton's equations can be
analyzed by the qualitative investigations and by the numerical integration.
It is found that closely to dislocation and symmetrically about it the two
bounded domains exist, in which the effective mass of transversal motion is
negative. Therefore the motion of a particle in these domain must be finite.
The motion integrals other than energy supposedly don't exist, and because
of this the quasi-classical quantization can't be performed.

\subsubsection{\qquad3.5. One-Electron States in Deformed Crystal in
Magnetic Field}

One-electron states in a deformed crystal in the magnetic field were
researched by the quasi-classical quantization the motion of the
corresponding classical particle. If the hamiltonian parameters to a good
approximation can be supposed to be constant in the area that contains the
electron trajectory, this trajectory is a helical line with the axis
directed along the magnetic field. The projection of this trajectory on the
plane perpendicular to the magnetic field is an ellipse, and this motion can
be quantized by the Bohr - Sommerfeld quantization rules with the discrete
equidistant energy spectrum as a result. As usually (see the monograph
[31]), there is supposed in the section [1.3.6] that the energetic spectrum
can be described as the superposition of the equidistant sequences, zero
points of which are determined by local values of the potential energy, if
the potential is a smooth function of coordinates, so that it is little
varied within the limits of the Larmor's orbits under considered energies.
More recently (unpublished) the author has concluded that this supposition
that means the electron stationary state to be localized close to the
corresponding classical orbit is incorrect. The Larmor's radius determines
the wave function localization only in a direction perpendicular to some
line that can have a length of the order of the dimension of the all
potential well, in which the electron is situated. This line is determined
by the choice of the coordinate system and its origin, and therefore this
localization has a physical sense only, when the space inhomogeneity
determines this choice. In the case of the solving in the rectangular
coordinates, \textit{e. g.,} (see the monograph [32]) there are straight
lines $y=y_{0},$ where $y_{0}$ are the eigenvalues of the orbit center
operator that has a discrete spectrum. In so doing the orbit center abscissa
remains to be uncertain. When the solution is taken in the polar coordinate
system, the state is characterized by the orbit center distance from the
coordinates origin that can take the discrete values, but the center azimuth
remains to be uncertain. This uncertainty is peculiar to this problem,
because the operators of the coordinates of the orbit center commutate with
the hamiltonian, but don't commutate with one another. The solutions of the
time Schr\"{o}dinger equation that are the Gaussian wave packets, which move
at the fixed in the space classical orbits with the minimally uncertain
centers, what are called as coherent states, are not the hamiltonian
eigenfunctions. The author has obtained the exact solution of the
quantum-mechanical problem about an electron in the uniform magnetic field
and the step potential (unpublished) and showed that this energy spectrum
doesn't coincide with one obtained by the quantization of the motion
integral of the corresponding classical problem. In so doing the largest
differences are obtained at large quantum numbers. Therefore the question
about the thermodynamic and galvanomagnetic characteristics of the electron
gas in the nonuniform potential field requires further investigation.

It appears to be evident that the trajectory of the classical particle, the
hamiltonian of which is obtained from the effective hamiltonian with the
uniform magnetic field, must be unlocked similar to the Burgers contour,
when this particle envelopes the dislocation line. In the section [1.3.6] it
is justified by the solving of the corresponding motion equations. In the
case of the screw dislocation parallel to the magnetic field this doesn't
break the locking of the trajectory projection on the plane perpendicular to
the field, and the motion integrals: the angular momentum with respect to
the dislocation line, and the radial component of the action - can be
quantized. If the dislocation core radius would be considered zero, the
transversal motion hamiltonian coincide with the electron hamiltonian in the
uniform magnetic field, and the field of the magnetic string, \textit{i. e. }%
infinitely thin solenoid, the flux of which is proportional to the Burgers
vector and to the longitudinal momentum. The exact solution of the
corresponding quantum-mechanical problem is known (see the work [33]). Its
spectrum coincides with one that obtained by the quasi-classical
quantization. In the case of several screw dislocations or magnetic strings
the axial symmetry is destroyed, but the classical problem retains to be
solvable for the trajectories, which don't cross the dislocation lines. On
the base of this solution a hypothesis about the electron energy spectrum in
this field was suggested in the section [1.3.6]. But recently (unpublished)
the author has obtained the exact solution of this quantum-mechanical
problem, from which follows that this hypothesis is incorrect.This problem
also requires further investigation.4. Many-Electron Phenomena in Deformed
Crystal

\subsubsection{\qquad4.1. Electrical Field in Deformed Conductors}

When the effective hamiltonian was derived in the section [1.2.4], it was
supposed that the effect of the lattice atoms on the electron goes to zero
at a distance of several lattice constants. If other electrons are in the
crystal, and all or some part of atoms are ions, they generate the electric
field that can't be supposed proportional to the deformation in the point
under consideration. In the section [1.4.1] in the self-consistent field
approximation the equation for the electrostatic potential generated by the
all electrons and ions of the deformed crystal is derived. This potential
must be included into the one-electron effective hamiltonian.

Let us describe shortly this derivation. The smooth component of the
nonuniform density of the ion charge is proportional to dilatation, \textit{%
i. e. }the trace of the distortion tensor. The nonuniform electron density
is formed self-consistently, \textit{i. e.} so that the flow, which should
be generated by the electric field, is entirely compensated by the diffusion
flow generated by the electron density heterogeneity. The total electric
charge density is a smooth function and generates a smooth self-consistent
electric field. The potential of this field can be taken away from the
potential energy of the one-electron Schr\"{o}dinger equation. Then the
residual is the short-range ion potential that was considered in the section
[1.2.4]. It is not periodic in the deformed crystal. If in the Lagrange's
curvilinear coordinates the periodic term would be detached, the residual
term will be small, because it is proportional to distortion, but non-smooth
function of coordinates. This residual term leads to the smooth deformation
potential in the effective hamiltonian. The deducted potential of the
self-consistent field as the smooth function can be added to the potential
energy of this hamiltonian. Then the electron density amendment depending on
the potential of this self-consistent field would be obtained by the
perturbation theory. On the other hand, this potential is related with the
density of electrons and ions by the Poisson's equation. As a result the
closed equation for the self-consistent field potential would be obtained.
Generally it is integro-differential equation. In the case, \textit{e. g. },
of non-degenerated electron gas it has a form:

$\ \ \ \ \ \ \ \ \ \ \ \ \ \ \ \ \ \ \ \ \ \ \ \Delta\varphi-\lambda
^{2}\varphi+\frac{\lambda^{2}}{V}\int\limits_{V}\varphi dV=-\frac
{1}{\varepsilon_{0}}\Xi\left( \mathbf{r}\right) +\frac{1}{\varepsilon_{0}V}%
\int\limits_{V}\Xi\left( \mathbf{r}\right) dV-\frac{q}{\varepsilon_{0}}.$ \
\ \ \ \ \ \ \ \ \ \ \ \ \ \ \ \ \ \ \ \ \ \ \ \ \ \ \ \ \ \ \ \ \ \ \ \ \ \
\ \ \ \ Here

$\ \ \ \ \ \ \ \ \ \ \ \ \ \ \ \ \ \ \ \ \ \ \ \ \ \ \ \ \ \ \ \ \Xi\left(
\mathbf{r}\right) =en_{0}\left[ \frac{1}{2}\mu_{ij}T_{klji}w_{kl}+\frac
{U}{kT}+w_{ll}\left( \mathbf{r}\right) \right] $ \ \ \ \ \ \ \ \ \ \ \ \ \ \
\ \ \ \ \ \ \ \ \ \ \ \ \ \ \ \ \ \ \ \ \ \ \ \ \ \ \ \ \ \ \ \ \ \ \ \ \ \
\ \ \ \ \ \ \ \ \ \ \ \ \ \ \ \ \ is the inoculating heterogeneity of the
charge density that is given by the trace of the tensor of the reciprocal
effective mass (see formula (2)), the deformation potential $U,$ and the
heterogeneity of the positive charge density proportional to the dilatation;
$\lambda$ is the reciprocal of the shielding radius; $V$ is the crystal
volume; $q$ is the crystal charge; $e$ is the modulus of the electron
charge; $n_{0}$ is the electron density in the non-deformed crystal. The
solution of this equation must be got in the all volume the conductor and be
joined on the bound with the solution of the Poisson's equation for the
surroundings. In metals, \textit{i. e. }in a conductor with degenerated
electron gas, instead of the multiplying by $\lambda^{2}$ the integral
operator with the core defined by metal electron characteristics appears. In
the case of a point inoculating charge it leads to the well-known shielding
charge density oscillations. These equations in the conducting bodies play
the same role as the Poisson's equation in the non-conducting ones. They
form a new method in the electrostatic of continuous mediums, in which the
electric field in the conductor isn't supposed to be zero a priori. The
solution can always be found as the sum of the partial solution of the
nonuniform equation with zero boundary conditions and the general solution
of the uniform equation containing the voluntary constants, with help of
which the boundary conditions can be satisfied. As an example, the problem
is considered about the electric field within the thickness of the hollow
cylinder and beyond it, when the cylinder is deformed as if an edge
dislocation with the macroscopic Burgers vector were situated on its axis.
The solution shows that this cylinder generates the field similar to the
dipole field, which can be not very small. The equations can also be used
for the uniform, but bounded conductor with $\Xi\left( \mathbf{r}\right) =0$
in an external field. For example, in the case of a conductor with the
non-degenerated electron gas the solution of the corresponding equation
describes the exponential decreasing of the external field with moving from
the bound into the thickness. The approximate solution of the problem about
an uniform plate with the degenerated electron gas in the uniform external
field perpendicular to it shows that the field in the plate aside from a
term exponentially decreasing with moving off the surface contains a small
oscillating amendment that has the maximum in the center of the plate.

\subsubsection{\qquad4.2. Heterogeneous Kinetic Coefficient in Deformed
Conductor}

In the next section [1.4.2] the transport phenomena in the deformed
conductor are studied. In the theory of the transport phenomena an incorrect
method was in general use, in which dislocations were considered similarly
to point scattering centers. In fact it is obvious that the scattering by
the long-range fields can't be accounted by the Boltzmann integral. It is
incorrect also to consider the scattering by several dislocations as the sum
of the effects from single dislocations. In most important cases the method
by S. I. Pecar is applicable, in which it is supposed that other scatterers
cause so small distance of the electron free path that the deformation
generated by the dislocation can be considered on this distance as uniform.
Then let us determine the kinetic coefficients in any uniformly deformed
volume with the dimensions much larger than the distance of the electron
free path, and next solve the problems for the non-uniformly conducting
bodies. In the works using this method the symmetrical case, when the
kinetic coefficients are scalars, was commonly considered. Furthermore,
there was taken into account only variation of the electron density that
satisfies to the quasi-neutral requirement. As result in the deformed
conductor the conductivity also would remain scalar, though it would depend
on coordinates. In the section [1.4.2] the anisotropy of the effective mass
generated by the uniform shear deformation is also taken into account. By
solving the kinetic equation the tensor kinetic coefficients are obtained
that can explain some experimental effects. As an example the flowing of the
electric current parallel to the screw dislocation must be accompanied by
the flowing of the circular currents that envelops its axis, and
consequently by the axial component of the magnetic field.

\subsubsection{4.3. Transition to Superconducting State and Current in
Deformed Superconductors}

The influence of the deformation on various phenomena in superconductors is
considered in the sections [1.4.3], and [1.4.4] on the base of plausible or
experimentally confirmed phenomenological suppositions. To cite an example,
on the base of the known fact of the dependence of the superconducting
transition temperature on the lattice hydrostatic compression, a supposition
is made that in the superconductor with smoothly non-uniform dilatation a
field of the local transition temperature $T_{C}\left( \mathbf{r}\right) $
can be introduced. We can consider this field as smooth, if the correlation
radius of the order parameter fluctuations determined by the modulus of the
difference $\left| T-T_{C}\left( \mathbf{r}\right) \right| $ is (at various
temperatures in the various areas) much smaller than the characteristic
distance of the field variation. Then in the areas in which the difference $%
T-T_{C}\left( \mathbf{r}\right) <0$ and is sufficiently large modulo the
ordered regions should appear. The appearance of this region close to an
edge dislocation predicted firstly in the work \ [34] is studied in more
detail in the section [1.4.3]. The appearance of the ordered region close to
a screw dislocation and the influence of the magnetic field on this
phenomenon is also described. The general theory of phase transitions of the
second kind in systems heterogeneous in large scale, which was developed in
works by M. A. Krivoglaz and the author (see the review [35]) holds that the
appearance of these ordered regions isn't the true phase transition. The
ordered regions would form a framework piercing through all the
superconductor, but they are broken up into domains with random values of
the order parameter phase by zones of the phase slip. In these zones the
modulus of the order parameter approaches to zero and thereafter increases
anew to the equilibrium value.\textbf{\ }The true superconductivity in this
framework shall also be absent so long as the averaged domain length will be
smaller than the averaged length of the superconducting regions between the
points of their intersections with other ones. Only when this condition will
be satisfied, it will be possible to pass over the framework of the
superconducting regions at any distance without intersection of any boundary
of the phase slip, \textit{i. e. }in the system the long-range order will
appear.

Some effects, which follow from the supposition that the deformation of the
superconductor can make it anisotropic, are considered. In so doing the
proportionality coefficient $\mathbb{P}$ in the equation for the
superconduction current density

$\ \ \ \ \ \ \ \ \ \ \ \ \ \ \ \ \ \ \ \ \ \ \ \ \ \ \ \ \ \ \ \ \ \ \ \ \ \
\ \ \ \ \ \ \ \ \ \ \ \ \ \ \ \ \ \ \ \ \mathbf{j=}\mathbb{P}en_{s}\left(
\hbar\nabla\Phi-\frac{2e}{c}\mathbf{A}\right) $ \ \ \ \ \ \ \ \ \ \ \ \ \ \
\ \ \ \ \ \ \ \ \ \ \ \ \ \ \ \ \ \ \ \ \ \ \ \ \ \ \ \ becomes a tensor of
the second rank that depends on coordinates. Here $n_{s}$ is the density of
the superconducting electrons, $\Phi$ is the order parameter phase, $\mathbf{%
A}$\textbf{\ }is the magnetic field vector potential. After eliminating $%
\nabla\Phi$ with the help of the operation ''curl'' and $\mathbf{A}$ by the
use of the Maxwell's equations the closed differential equation for $\mathbf{%
j}$ is obtained. Its solution in the case, when $\mathbb{P}$ depends
linearly on the deformation generated in the hollow cylinder by the screw
dislocation at its axis describes the paradoxical effect: the longitudinal
transport current parallel to the screw dislocation is accompanied by the
circular current that envelopes the dislocation . This current flows close
to the outer surface of the cylinder and generates the longitudinal magnetic
field, which increases deep into the superconductor and passes a maximum at
the distance the London's length.

\subsubsection{\qquad4.4. Abrikosov's Vortex in Anisotropic and Deformed
Superconductor}

The deformation amendments to the tensor $\mathbb{P}$ and the
superconducting transition temperature cause the interaction of an
Abrikosov's vortex with the deformation field. If the deformation varies on
the distance, which is larger than the London's length the studying of its
effect on the Abrikosov's vortex can be reduced to the consideration of the
energy of this vortex fragment in a uniformly deformed superconductor. The
high-$T_{C}$-superconductors, as a rule, are highly anisotropic in
non-deformed condition as well. Therefore the characteristics of the
Abrikosov's vortex in the anisotropic superconductor are very interesting.

The behavior of a straight Abrikosov's vortex in an anisotropic uniaxial
London's superconductor is studied in [1.4.4]. Analytical expressions are
derived that approximately describe the magnetic field in three regions: the
asymptotic region, where the distance $r$ from the vortex line is greater
than $\lambda\Gamma$ ($\lambda$ is the London's length and $\Gamma$ is the
anisotropy constant), the intermediate region $\lambda<r<\lambda\Gamma$, and
the region $r<\lambda$. It is found that, when the anisotropy is large, in
the intermediate region the component of the magnetic field along the vortex
line changes its sign for a certain interval of angles between the vortex
line and the anisotropy axis. Because of this the interaction of parallel
vortices, whose plane is parallel to the anisotropy axis, has a minimum and
a maximum. This means that numerous metastable vortex lattices can exist.
Additional terms in the vortex self-energy are obtained, and although they
are smaller than the leading logarithmic term, they display a different
dependence on the angle between the vortex line and the anisotropy axis.
Furthermore in the section [1.4.4] it is shown that the common opinion about
increasing the Abrikosov's vortex specific energy in bending is incorrect.
Because of attraction of the parallel currents forming an Abrikosov's vortex
the vortex energy per unit length decreases, while bending the vortex, by a
quantity proportional to the square of the curvature. Solving the London's
equation in an approximation allowing for this effect makes it possible to
calculate the energy of an Abrikosov's vortex, which has the form of a
helix, whose radius and pitch are much larger than the correlation length,
whose curvature is small compared to the reciprocal London's length, and
whose slope in relation to an axis coinciding with the direction, in which
vortex energy is the highest, is also small. When the anisotropy is large,
which is characteristic of high-$T_{C}$ -superconductors, the energy of such
Abrikosov's vortex is lower than of a straight Abrikosov's vortex. Certain
consequences of the fact that the Abrikosov's vortices in a high-$T_{C}$
-superconductor are helical are discussed. Among these is a phase transition
that breaks symmetry between Abrikosov's vortices shaped like right- and
left-hand helixes in relation to the magnetic field.

The interaction of Abrikosov's vortex with dislocations perpendicular to
Cu-O layers in high-$T_{C}$ -superconductor at distances greater than the
characteristic dimensions of the vortex and of the dislocation is discussed.
This interaction is caused by local changes of the order parameter with the
dilatation, as well as of the direction of the anisotropy axis with the
shear deformation. It is shown that such interaction does not produce any
centers of pinning, but facilitates the attraction of a vortex to the
dislocation core. And if a thermal fluctuation will throw the vortex over
the dislocation core, this interaction will repel the vortex from the
dislocation. Coiling of the vortex around a screw dislocation core in the
direction of helicoid formed by Cu-O layers in a crystal with a dislocation
can be a pinning center.

\subsubsection{\qquad\textbf{Acknowledgments}}

In summary I would express a sincere gratitude to the scientists from
theoretic departments of the Institute for Physics of Metals of National
Academy of Sciences of Ukraine and especially to professor Michail Krivoglaz
for many useful discussions in the course of executing works which
constituted the monograph [1] and this article, and to my wife, doctor of
physical and mathematical science D.V.Lotsko for great help in preparing
manuscripts.

\subsection{\textbf{References}}

[1] I. M. Dubrovskii, \textit{Theory of Electron Phenomena in Deformed
Crystal }(RIO IMF, Kyiv, 1999). [in Russian].

[2] S. I. Pecar, \textit{Investigations into Electron Theory of Crystals }%
(Gosizdat teckniko-teoreticheskoj literatury, Moscow-Leningrad, 1951). [in
Russian].

[3] J. M. Luttinger and W. Kohn, Phys. Rev. \textbf{\ 97, }869 (1955).

[4]\ J. M. Luttinger, Phys.Rev. \textbf{102}, 1030 (1956).

[5] J. Bardeen and W. Shokley, Phys. Rev. \textbf{80, }72 (1950).

[6] M. F. Dejgen and S. I. Pecar, Zh. Eksp. Teor. Fiz. \textbf{21}, 803
(1951). [in Russian].

[7] S.C. Hanter and R.H. Nabarro, Proc.Roy.Soc. \textbf{A220}, 542 (1953).

[8] A. I. Ahiezer, M. I. Kaganov and G. J. L'ubarskii, Zh. Eksp. Teor. Fiz.%
\textbf{\ 32}, 837 (1957).[in Russian].

[9] V. M. Kontorovich, Zh. Eksp. Teor. Fiz. \textbf{45}, 1638 (1963).[in
Russian].

[10] V. M. Kontorovich, Usp. Fiz. Nauk. \textbf{142}, 265 (1984). [in
Russian].

[11] J.Th.M.De Hosson, H.P.Van De Braak and W. J. Caspers, Phys. Lett.%
\textbf{\ A63}, 174 (1977).

[12] S. Winter, phys. stat. sol. \textbf{B79}, 637 (1977) .

[13] S. Winter, phys. stat. sol. \textbf{B90}, 101 (1978).

[14] V. L. Gurevich, I. P. Lang and S. T. Pavlov, Zh. Eksp. Teor. Fiz.
\textbf{59}, 1679 (1970). [in Russian].

[15] I. P. Lang and S. T. Pavlov, Fiz. Tverd. Tela. \textbf{12}, 2412
(1970). [in Russian].

[16] G. D. Whitfield, Phys.Rev. \textbf{121}, 720 (1961).

[17] G. L. Bir and G. E. Pikus, \textit{Symmetry and Deformational Effect in
Semiconductors }(Nauka, Moscow, 1972). [in Russian].

[18] S.V. Iordanskii and A. E. Koshelev, Zh. Eksp. Teor. Fiz. \textbf{90},
1399 (1986). [in Russian].

[19] K. Z. Kawamura, Z.Physik, \textbf{B29,} 101 (1978).

[20] R. A. Brown, J.Phys.F. \textbf{9}, L241 (1979).

[21] H. Teichler, Phys.Lett. \textbf{A87, }113 (1981).

[22] R. A. Brown, Australian Journal of Physics. \textbf{36}, 321 (1983).

[23] R. A. Brown and M. A. Oldfield, J.Phys.C. \textbf{21}, 4437 (1988).

[24] P. B. Hirsch, A. Howie, R. B. Nicholson, D.W. Pashley and M. J. Whelan,
\textit{Electron Microscopy of Thin Crystals }(Butter worths, London, 1965).

[25] S. Takagi, J. Phys. Soc. Jap. \textbf{26}, 1239 (1969).

[26] M. A. Krivoglaz, \textit{X-Ray and Neutron Diffraction in Nonideal
Crystals }(Springer-Verlag, Berlin, Heidelberg, 1996).

[27] I. M. Lifshits, Usp. Fiz. Nauk. \textbf{83, }617 (1964).

[28] H. Stehle and A. Seeger, Zs.Phys. \textbf{146}, 217 (1956).

[29] H Stehle and A. Seeger, Zs.Phys. \textbf{146}, 242 (1956).

[30] N. Fr\"{o}man and P. O. Fr\"{o}man,\textit{\ JWKB Approximation.
Contribution to the Theory (}North-Holland Publishing Company, Amsterdam,
1965).

[31] I. M. Lifshits, M. J. Azbel and M. J. Kaganov, \textit{Electron Theory
of Metals} (Nauka, Moscow, 1971).\textit{\ } [in Russian].

[32] L. D. Landau and E. M. Lifshits, \textit{Quantum Mechanics.
Nonrelativistic Theory} (Fizmatgiz, Moscow, 1963). [in Russian].

[33] S.V. Iordanskii and A. E. Koshelev, Zh. Eksp. Teor. Fiz. \textbf{91},
326 (1986). [in Russian].

[34] V. M. Nabutovskii and B. J. Shapiro, Pisma Zh. Eksp. Teor. Fiz. \textbf{%
26}, 624 (1977). [in Russian].

[35] I. M. Dubrovskii, in \textit{Fizika realnyh kristalov }(Naukova dumka,
Kyiv, 1992), pp. 150 - 168, and references therein. [in Russian].

\bigskip

\end{document}